\documentstyle[preprint,prb,aps,tighten]{revtex}

\begin{document}

\draft
 
\title{Beyond Gross-Pitaevskii: local density vs. correlated basis approach 
for trapped bosons} 

\author{A. Fabrocini$^{1)}$ and A.Polls$^{2)}$}
\address{
$^1)$ Dipartimento di Fisica, Universit\`a di Pisa, 
I-56100 Pisa, Italy \protect\\
$^2)$ 
Departament d'Estructura i Constituents de la Mat\`eria, Universitat de Barcelona,
 E-08028 Barcelona, Spain}

\maketitle

\begin{abstract}

We study the ground state of a system of Bose hard-spheres trapped in 
an isotropic  harmonic potential to investigate the effect of the 
interatomic correlations and the accuracy of the Gross-Pitaevskii equation.
We compare a local density approximation, based on the energy functional 
derived from the low density expansion of the energy of the uniform 
hard sphere gas, and a correlated wave function approach which explicitly 
introduces the correlations induced by the potential. Both higher order terms 
in the low density expansion, beyond Gross-Pitaevskii, and explicit 
dynamical correlations have effects of the order of percent when the number 
of trapped particles becomes similar to that attained in recent experiments 
(N$\sim 10^7$). 
\end{abstract}


\narrowtext

The recent discovery of Bose-Einstein condensation (BEC) of magnetically
trapped alkali atoms has generated a huge amount of theoretical investigations. 
A review of the present situation can be found in 
refs.\cite{Dalfovo98,Parkins98}.
Present experimental conditions are such that the atomic gas is very dilute,
i.e., the average distance among the atoms is much larger than the range of
 the interaction. As a consequence, 
the physics should be dominated by two body collisions, 
generally well described in terms of the $s$-wave scattering length.
The case of a positive scattering length is equivalent to consider a very
 dilute system of hard spheres (HS), whose diameter coincides with the 
scattering length itself. So, the Gross-Pitaevskii (GP) theory for weakly
interacting bosons seems the logical tool to study  these systems  and 
most of the present days theoretical work is founded on (or has its starting 
point in) this theory \cite{Pita61}. However, in very recent experiments 
the number of trapped atoms has spectacularly increased reaching N values 
 of the order $10^6$ and $10^7$ atoms\cite{Stamper98}. 
Therefore, it seems logical to ask for a deeper study of the effect of 
the interatomic correlations and of the accuracy of the GP scheme in 
this new scenario. 

In order to address these questions, we  study the ground state of  a 
system of Bose hard spheres trapped by a harmonic potential. 
More precisely, we consider hard spheres with a diameter
of $52.9 \AA$, corresponding  to the $s$-wave triplet-spin 
scattering length of $^{87}$Rb, in an isotropic harmonic trap 
characterized by an angular frequency $\omega/2 \pi = 77.78 $Hz. 
We also examine the large N atomic sodium case of ref.\cite{Stamper98}. 
We use and compare two methods: {\em (i)} a local density approximation 
(LDA) based on an energy functional derived by the low--density 
expansion of the energy of an uniform hard sphere gas; {\em (ii)} a 
correlated basis function (CBF) approach which explicitely takes 
into account the dynamical correlations induced by the potential and 
 which is not, in principle, limited to purely repulsive interactions.

 {\em \underline { LDA theory.}}  
In the case of an uniform hard sphere  Bose gas, a 
perturbation theory in the expansion parameter $x=n a^3$, 
where $n$ is the density of the system,
leads to the following low density (LD) expansion for the energy density 
\cite{Fetter71}:
\begin{equation}
\frac {E}{V}= \frac {2 \pi n^2 a \hbar^2 }{m} \left [ 
1+ \frac{128}{15} \left ( \frac
{na^3}{\pi} \right )^{1/2} + 8 (\frac {4}{3} \pi - \sqrt {3} ) (na^3) \ln (na^3)
 + O(na^3) \right ].
\label{eq:HHP}
\end {equation}

The energy functional associated with the GP theory  
is simply obtained in LDA by  keeping only the first term in  the 
expansion (\ref{eq:HHP}):  
\begin{equation}
E_{GP}[\psi] = \int d{\bf r} \left [ \frac {\hbar^2}{2 m} 
\mid \nabla \psi(r)\mid ^2 +
\frac {m}{2} \omega^2 r^2  \mid \psi \mid ^2+ \frac {2 \pi \hbar^2 a}{m} 
\mid \psi\mid ^4
\right ],
\label{eq:GPE}
\end{equation}
where the $wave function$ $\psi$ is normalized to the 
total number of atoms.

By performing a functional variation of $E_{GP}[\psi]$ 
one finds the Euler-Lagrange equation, known as the GP equation
\begin{equation}
 \left [ - \frac {\hbar^2}{2 m} \nabla ^2 +
\frac {m}{2} \omega^2 r^2  + \frac {4 \pi \hbar^2 a }{m} \mid \psi\mid ^2
\right ] \psi = \mu \psi ,
\label{eq:GPMU}
\end{equation}
where $\mu$ is the chemical potential. This equation has the form of 
a nonlinear stationary Schr\"odinger equation, and it has been solved for 
several types of traps using different numerical methods 
\cite{Edwards95,Dalfovo96,Cap98,Arimondo98}.

A logical step further, in the spirit of LDA, is to include into 
the energy  functional  the next terms of the correlation energy  
(\ref{eq:HHP}). It is convenient to simplify the
notation by expressing lengths and  energies in harmonic oscillator (HO) 
units. The HO length is  $a_{HO} = ( \hbar/m\omega)^{1/2}$,
and the spatial coordinates, the energy and the wave function are rescaled as 
${\bf r}= a_{HO}{} \bar {\bf r}$, $E= \hbar\omega E_1$ and  
$\psi(r)= (N/a^3_{HO})^{1/2} \psi_1(\bar r)$,
 where $\psi_1(\bar r)$ is normalized to unity.

By taking into account the next terms of the correlation energy, LDA 
provides a modified GP (MGP) energy functional 
\begin{eqnarray}
 E_{MGP}[\psi_1] & = & \int d\bar {\bf r} \left [\frac {1}{2} 
 \mid \nabla_{\bar r} \psi_1(\bar r)\mid ^2 + \frac {1}{2} 
\bar r^2 \mid \psi_1(\bar r)\mid ^2
 + 2 \pi \bar a N \mid \Psi_1(\bar r)\mid ^4 \right. \nonumber \\
& &  \left. \left ( 1 + \frac {128}{15} (N \bar a^3 \mid \psi_1(\bar r)
\mid ^2 )^{1/2}+8 (\frac {4 \pi}{3} -\sqrt{3}) N \bar a^3 
\mid \psi_1(\bar r)\mid ^2
 \ln {(
N \bar a^3 \mid \psi_1(\bar r)\mid^2)} \right ) \right ],
\end{eqnarray}
and  a  modified Gross-Pitaevskii equation 
\begin{eqnarray}
 \left [- \frac{1}{2} \nabla_{\bar r}^2 \right. & + & \left. 
\frac{1}{2} \bar r^2 +
 4 \pi \bar a N \mid \psi_1(\bar r) \mid^2 
 + 5 \pi \bar a^{5/2} N^{3/2} \frac {128}{15} 
\mid \psi_1(\bar r)\mid ^3  \right. \nonumber \\
& & \left.
+ 8 \pi \bar a^4 N^2 (\frac {4 \pi}{3} -\sqrt{3} ) \mid 
\psi_(\bar r) \mid ^5 \{ 6 \ln {(N\bar a^3 \mid \psi_1(\bar r)\mid ^2) +2 }\} 
\right ] \psi_1(\bar r)= \mu_1 \psi_1(\bar r)
\end{eqnarray}
where $\bar a = a/a_{HO}$ and $\mu_1$ is the chemical potential 
in HO units. The MGP equation contains extra nonlinar terms in $\psi_1$.

{\em \underline  {CBF approach.}}  CBF theory is 
a powerful tool to study strongly interacting 
many--body systems (for a review, see ref.\cite{Valencia98}). 
In particular, it was used to study the HS homogenous Fermi gas 
in order to ascertain the accuracy of low density type 
expansions\cite{Fabrocini80}.
 
For N interacting bosons, at T=0 temperature, described by the 
hamiltonian
\begin{equation}
H= -\frac {\hbar^2}{2m} \sum_i \nabla^2_i+\sum_i V_{ext}({\bf r}_i)+
\sum_{i<j} V(r_{ij}),
\label{eq:hamiltonian}
\end{equation}
where $V_{ext}({\bf r}_i)$ is the confining potential and $V(r_{ij})$ is the 
interatomic potential, the CBF ground-state wave function is 
\begin{equation}
\psi_{CBF}(1,..,N)=F(1,..,N)\psi_{MF}(1,..,N).
\label{eq:wave_funct}
\end{equation}
where $F(1,..,N)$ is a many--body {\em correlation operator} applied 
  to the mean field w.f. $\psi_{MF}$.
 The advantage of using 
a CBF basis lies in the fact that non perturbative effects, 
as the short range repulsion for the hard spheres, 
may be directly inserted into the correlation. 

The simplest correlation operator has the Jastrow form\cite{Jastrow}
\begin{equation}
F(1,..,N)=\prod_{i<j} f_J(r_{ij}),
\label{eq:Jastrow_corr}
\end{equation}
where the Jastrow correlation function, $f_J(r)$, depends on the 
interparticle distance only. $f_J(r)$ is  variationally  determined 
 by minimizing the expectation value of 
$E_{CBF}=\langle \psi_{CBF} \vert H \vert \psi_{CBF} \rangle /
\langle \psi_{CBF} \vert \psi_{CBF} \rangle $. 
$E_{CBF}$ may be evaluated either by MonteCarlo techniques 
or by cluster expansions of the needed one-- and two--body densities, 
$\rho_1({\bf r}_1)$ and $\rho_2({\bf r}_1, {\bf r}_2)$. 

The energy per particle can be written as $E_{CBF}/N=T_{\rho}+V_e+V_{corr}$,  
where
\begin{equation}
T_{\rho}={\frac {1}{N}}{\frac {\hbar^2}{2m}}
\int d{\bf r}_1 [\vec \nabla_1 \rho_1^{1/2}({\bf r}_1)]^2,
\label{eq:trho}
\end{equation}
\begin{equation}
V_e={\frac {1}{N}}\int d{\bf r}_1\rho_1({\bf r}_1) V_{ext}({\bf r}_1),
\label{eq:ve}
\end{equation}
and the correlation energy, $V_{corr}$, is
\begin{eqnarray}
V_{corr} & = & 
{\frac {1}{N}}{\frac {1}{2}}
\int d{\bf r}_1\int d{\bf r}_2 \rho_2({\bf r}_1,{\bf r}_2)
[ V(r_{12})-{\frac {\hbar^2}{2m}} {\vec \nabla}^2 \ln f_J(r_{12})] \nonumber \\
& - &   {\frac {1}{N}}{\frac {\hbar^2}{2m}} 
\int d{\bf r}_1\int d{\bf r}_2\rho_2({\bf r}_1,{\bf r}_2)
{\vec \nabla} \ln f_J(r_{12})\cdot \vec \nabla \ln \rho_1^{1/2}({\bf r_1}),
\label{eq:vjf}
\end{eqnarray}
where $\rho_1$ is normalized to $N$.

$\rho_2$ may be calculated by cluster expansion in powers of 
the dynamical correlation, $h(r)=f_J^2(r)-1$, 
and the integral Hypernetted Chain (HNC) equations\cite{HNC87} may 
be used to sum infinite classes of cluster diagrams. 

Given the low-density of the trapped bosons system, it looks plausible to 
start by a lowest  order (LO) cluster expansion. In this approximation  
$\rho_2^{(LO)}({\bf r}_1,{\bf r}_2)=\rho_1({\bf r}_1)\rho_1({\bf r}_2)
f^2_J(r_{12})$
and 
\begin{equation}
V_{corr}^{(LO)}=
{\frac {1}{N}}{\frac {1}{2}}
\int d{\bf r}_1\int d{\bf r}_2
\rho_2^{(LO)}({\bf r}_1,{\bf r}_2) V_{JF}(r_{12})
\label{eq:vjfLO}
\end{equation}
where $V_{JF}(r)=V(r)+(\hbar^2/m) [{\vec \nabla} \ln f_J(r)]^2 $ is  
the Jackson--Feenberg potential. 

The minimization of $E_{CBF}^{(LO)}$ with respect to 
$\rho_1$ leads to the LO--correlated Hartree (CH/LO) equation 
\begin{equation}
 \left [ - \frac {\hbar^2}{2 m} \nabla ^2+V_{ext}({\bf r})+ 
\int d{\bf r_1}\rho_1({\bf r}_1) f^2_J(s)V_{JF}(s)
\right ] \rho_1^{1/2}({\bf r})= \mu \rho_1^{1/2}({\bf r}) 
\label{eq:CHE}
\end{equation}
with $s=\vert {\bf r} - {\bf r}_1\vert$.  


The optimal choice for the Jastrow factor would be the one 
satisfying the Euler equation $\delta E_{CBF}/\delta f_J(r)=0$. 
Otherwise, parametrized functional forms may be chosen whose 
parameters are found through the minimization process. 
 We adopt here 
the correlation function minimizing the lowest order energy 
of a homogenous Bose gas with a healing condition at a distance 
$d$ (taken as variational parameter). For the HS case, 
 $f_J(r<a)=0$ and $f_J(r>a)=u(r)/r$ 
where $u(r)$ is solution of the Schr\"odinger--like 
equation $-u"=K^2u$.  $f_J(r)$ has the form\cite{PandaHS}
\begin{equation}
  f_J(r)={\frac {d}{r}}{\frac {\sin[K(r-a)]}{\sin[K(d-a)]}} 
\label{eq:correlation}
\end{equation}
where the healing conditions, $f_J(r\ge d)=1$ and 
$f'_J(r=d)=0$, are fixed by the relation: $\cot [K(d-a)]=(Kd)^{-1}$. 

 {\em \underline  {Results.}}  
 We begin  by briefly discussing the boson HS homogeneous case. It was shown 
in ref.\cite{PandaHS} that the lowest order cluster energy of an 
infinite system of bosonic hard spheres, correlated by the  
Jastrow factor (\ref{eq:correlation}), asymptotically tends to 
the first term of the LD expansion (LD0) in (\ref{eq:HHP})  
when the healing distance goes to infinity. 
We have numerically checked this behavior. The situation changes if 
all the many--body cluster terms are included via HNC summation, since 
the computed energy shows a clear minimum in $d$. Some results are given in   
 Fig.1 for various $x$--values. The energies have been multiplied by 
 $10^{3(2,1)}$ at $x=10^{-6(-5,-4)}$, respectively. The Figure shows the 
energy estimates computed by retaining different LD  expansion 
terms (LD1 and LD2 correspond to adding the second and third terms in 
(\ref{eq:HHP})), the HNC energies and, at the highest $x$--value, the exact 
 energy, evaluated by a diffusion MonteCarlo method\cite{Boronat98}. The 
quality of the HNC results is excellent as they practically coincide with 
the exact ones at all the densities (the low density value at low $x$ and 
the DMC one at large $x$). The convergence of the LD expansion at large 
$x$ appears to be rather poor, pointing to the relevance of successive 
contributions.  

The  GP, MGP and CH/LO equations for $\psi_1$  have been solved  by the 
steepest descent method\cite{Steepest} for the isotropic harmonic 
oscillator trap previously described. An initial trial state is projected 
onto the minimum of the functional by propagating 
it in imaginary time. In practice, one chooses an arbitrary time step 
$\Delta t$ and iterates the equation 
\begin{equation}
\psi_1({\bf r}_1,t+\Delta t) \approx \psi_1({\bf r}_1,t) - 
\Delta t H \psi_1({\bf r}_1,t) 
\end{equation}
by normalizing $\psi_1$ to 1 at each iteration. When the number of particles 
becomes very large the time step, which governs the rate of convergence, 
should be taken accordingly small. The number
of iterations substantially increases and it is convenient to start the
 convergence process from a reasonable wave function (for the GP and MGP 
equations we start from the ground state harmonic oscillator function which
minimizes the GP energy functional, while, in the CH/LO case, we start from 
the GP solution).              

We find that the LO approximation in the finite system shows a behavior 
similar to the infinite case. In particular, the solution of the CHE/LO 
equation provides an energy (in reduced unities) very close to the 
GP one when the healing distance becomes very large. In fact, for 
$N=10^5$ and $\bar a =4.33\times 10^{-3}$, corresponding to the $^{87}$Rb 
scattering length, we obtain $E_1^{(LO)}/N=12.57$, $12.28$ and $12.11$ at  
$\bar d=10$, $12$ and $15$, respectively, while  $E_1^{(GP)}/N=12.10$. 
This indicates that many--body effects can be recovered only by a 
HNC treatment also for a finite number of atoms. However, the solution 
of the fully correlated HNC Hartree equation, even if feasible in principle, is 
very cumbersome, computationally time consuming and numerically delicate. 
For this reason we have decided to estimate these effects by adopting a 
LDA--type approach to $V_{corr}$. 
We approximate $V_{corr}\sim V_{corr}^{LDA}$ where
\begin{equation}
V_{corr}^{LDA}={\frac {1}{N}} 
\int d{\bf r}_1\rho_1({\bf r}_1) E^{hom}_{HNC}(\rho_1).
\label{eq:corr_lda}
\end{equation}
$E^{hom}_{HNC}(\rho_1)$ is the HNC homogeneous gas energy per particle 
at density $\rho_1$. 

The minimization of the energy gives the HNC correlated Hartree 
equation (CH/HNC)
\begin{eqnarray}
 \left [- \frac{1}{2} \nabla_{\bar r}^2 \right. + \left. 
\frac{1}{2} \bar r^2 + E_{1,HNC}^{hom}(x) + 
x \frac {\partial  E_{1,HNC}^{hom}(x)}{\partial x}
\right ] \psi_1(\bar r)= \mu_1 \psi_1(\bar r)
\end{eqnarray}
where we have again introduced the scaled unities and 
$x=\rho_1 a^3=N\bar a^3 \vert \psi_1\vert ^2$.

The calculations have been performed for  the $^{87}$Rb scattering length. 
The scaled energies per particle and the root mean square radii  
  are  reported in Table I for 
particle numbers from $10^3$ to $10^7$. The Table also 
shows the results obtained by neglecting the kinetic energy term in the 
GP equation. This approach, lousely called Thomas
Fermi (TF) approximation,  has been discussed in the literature and allows 
for deriving simple analytical expressions \cite{Edwards95}. 
The differences between this TF approach and a rigorous one 
have been recently discussed \cite{Timmermans97,Schuck98} 
for spatially inhomogeneous Bose condensates. LDA has been 
used\cite{Dalfovo98} to estimate corrections to the GP energy 
within the TF approximation and retaining only the first correction 
in (\ref{eq:HHP}). The second correction 
is negative and partially cancels the first one. For instance,  
the cancellations go from $\sim 15\%$ for $N=10^4$ to $\sim 40\%$ 
at $N=10^6$ if we just take the central densities, whereas 
the final energy is reduced by  $\sim 15\%$ at $N=10^6$ and 
it is practically unaffected by the second correction at lower $N$--values.

As expected, the TF results are close  to the GP ones when N 
becomes large. The differences between GP and MGP increase with 
the number of particles and are of the order of $4\%$ for 
the chemical potential and $2.5 \%$ for the energy at $N=10^7$. 
The higher order terms in the low density expansion always have 
a repulsive effect. The same behavior is shown by the HNC results,  
 which, however, are less repulsive than MGP at the large $N$ values.

We notice that if one uses the GP solution to perturbatively estimate the 
MGP energy, then the correction 
 is negative (at $N=10^7$ , $\Delta E_1 = -4.54$). The non linear
character of the MGP equation is responsible for this discrepancy.

The density profile (normalized to unity) for $N=10^7$ particles is 
given in Fig.2.  For this large number of particles the TF and GP 
densities are close, 
whereas the more repulsive MGP and HNC solutions lower the
central density, expanding the density distribution and providing a larger 
radius, as shown in Table I . 

We have also considered a system of $N=1.5\times 10^7$ Na atoms ($a=27.5\,\AA$) 
in a spherical trap having a frequency of 230 Hz. These conditions roughly 
correspond to those of the experiment described in ref.\cite{Stamper98}.
The results are shown in the last row of the Table and in Fig.2. 
The effects of the correlations are similar to  those found in the large 
$N$ Rb cases. 
The energy increases by $\sim 1\%$  and the r.m.s. radius by $\sim 0.7\%$  
respect to GP. The HNC central density is slightly reduced. 


In conclusion, we find that both higher order terms in the low density 
expansion (beyond the GP approach and evaluated in LDA) and explicit 
dynamical correlations (induced by the strong repulsion) 
have a not negligible effect when the number of trapped particles 
becomes large. So, they must be properly considered. 

In this respect, CBF theory may play a preminent role. In addition, 
it  allows for a fully microscopic investigation for any 
type of potential. This is particularly 
interesting in view of its application to atoms having negative 
scattering lengths, as $^7$Li. In this case the potential exhibits  an 
actractive part and the GP equation has a metastable solution only 
if $N$ does not exceed some critical value $N_c$.  
A recent study\cite{Reatto98}, which makes use of an effective interaction, 
has shown that a new branch of Bose condensate may appear at higher densities. 
CBF may provide further insights into this problem having access to the 
full structure of the potential itself. 

The introduction of a HNC energy functional derived by the 
homogeneous HS system provides a quick and probably 
reliable way of embodying correlation effects into the  
treatment of the trapped atoms. We see as a particular 
appeal of this approach the clear possibilty of extending it to 
non spherical traps, as in real experiments. Moreover, as already stated, 
potentials other than the simple hard sphere one may be easily 
handled.


{\bf ACKNOWLEDGEMENTS}

One of us (A.F.) wants to thank Ennio Arimondo, Stefano Fantoni and 
Riccardo Mannella for several stimulating discussions. 
This research was partially supported by DGICYT (Spain) Grant
No. PB95-1249, the agreement CICYT (Spain)--INFN (Italy) and 
the Acci\'on Integrada program (Spain-Italy).

\begin{table}               
\bigskip
\caption{ 
Ground state energies per particle and chemical potentials of $N$ $^{87}$Rb 
atoms in an isotropic trap ($\omega/2\pi = 77.78\, Hz$). 
The last row refers to the Na case ($\omega/2\pi = 230\, Hz$).  
Energies in units of $\hbar\omega$ and radii in HO lengths.
}
\bigskip
\begin{tabular}{c| cccc |cccc|ccc}
\phantom{ca}&
\multicolumn{4}{c|}{$\mu_1$} & \multicolumn{4}{c}{E$_1$/N} & 
\multicolumn{3}{c}{$\sqrt{\langle r_1^2\rangle}$ } \cr 
\hline
 N & TF &  GP & MGP & HNC & TF & GP & MGP & HNC & GP & MGP & HNC \cr
\hline
10$^3$ \phantom{ca}&2.66  & 3.04 & 3.06 & 3.04 & 1.90 & 2.43 & 2.43 & 2.43 
&1.65&1.66&1.66\cr
10$^4$  \phantom{ca}&6.67  &6.87 & 6.92 &6.89 &6.87 &5.04 &5.08 & 5.04 
&2.44&2.45&2.44\cr
10$^5$  \phantom{ca}&16.75 &16.85 & 17.07 &16.94 &11.96 & 12.10 &12.25&12.20
&3.80&3.84&3.83\cr
10$^6$  \phantom{ca}&42.07 &42.12 & 42.97 &42.53 &30.05 & 30.12 &30.66&30.48
&6.01&6.10&6.06\cr
10$^7$  \phantom{ca}&105.68 &105.70 & 108.75&107.20&75.49&75.52&77.48&76.85
&9.52&9.74&9.64\cr
1.5$\times$10$^7$\phantom{ca}&91.07&91.10&92.41&91.67&65.05&65.09&65.92&65.66
&8.84&8.92&8.90\cr
\end{tabular}
\label{tab:gse}
\end{table}


\begin{figure}
\caption{ 
Energy per particle (in units of $\hbar^2/2ma^2$) for hogeneous hard spheres 
in function of $x$. See text.
} 
\label{fig:1}
\bigskip
\caption{ 
Density profiles for $N=10^7$ Rb atoms  
and for $N=1.5\times 10^7$ Na atoms  in different approaches 
(dotted line=GP, dashed line=MGP, solid line=HNC). 
Densities are normalized to unity and distances are in units of HO lengths.
} 
\label{fig:2}
\end{figure}
 
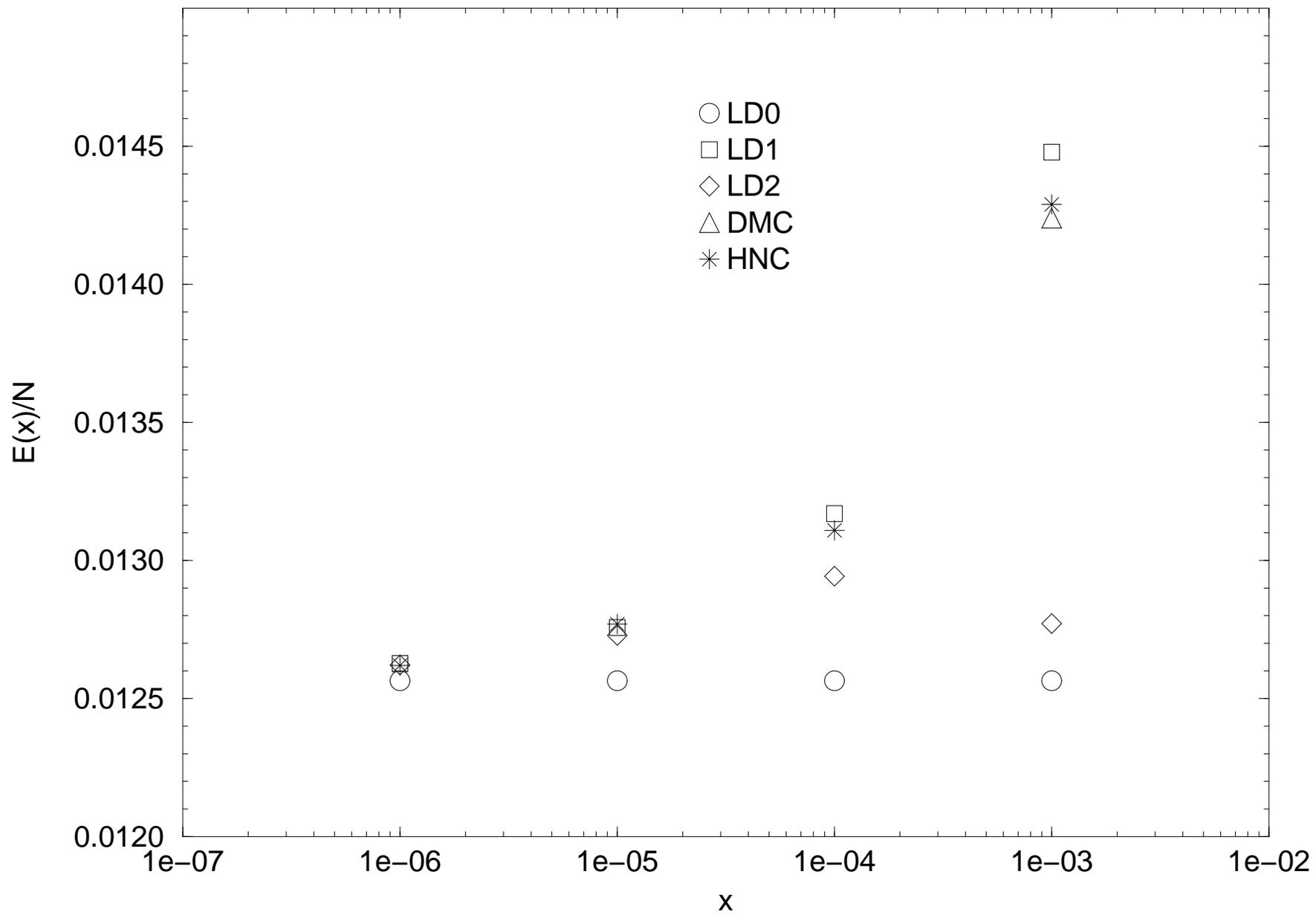
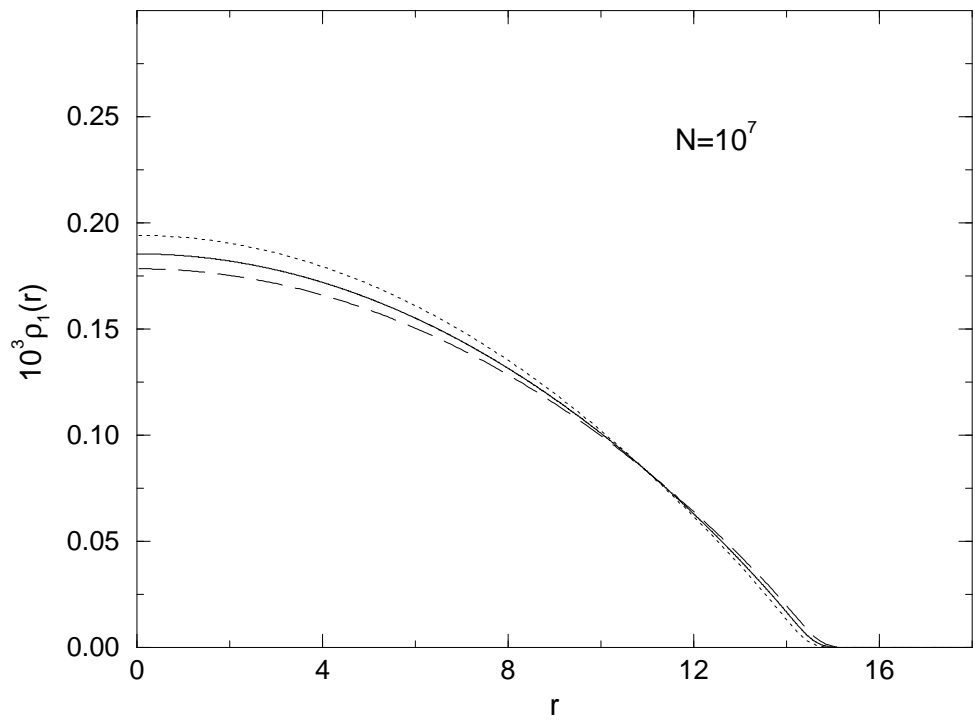
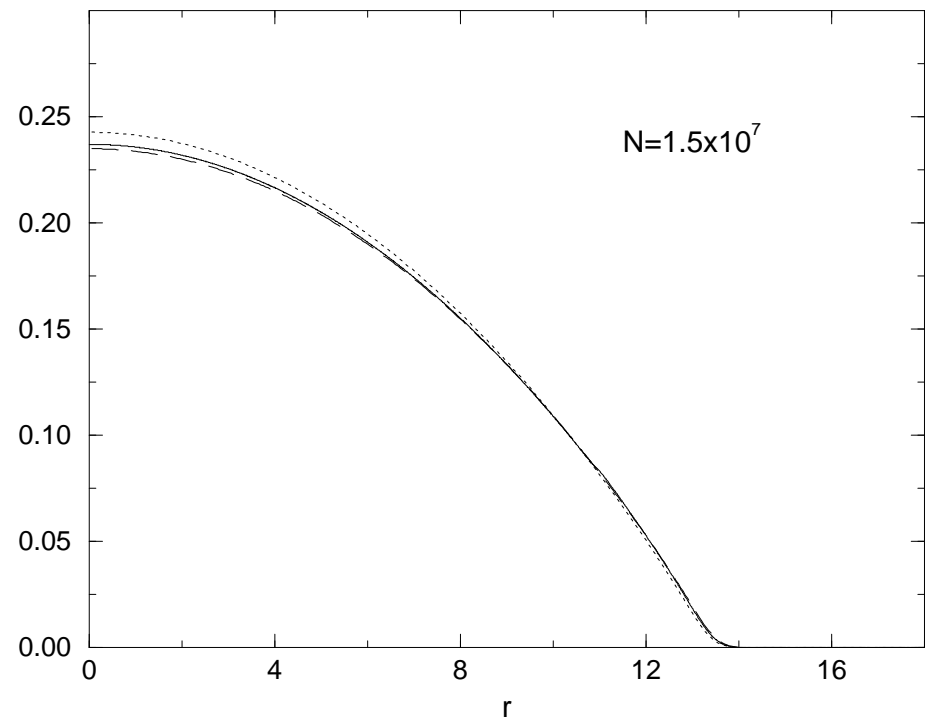
\end{document}